\documentclass[
  aps,
  prx,
  reprint,
  superscriptaddress,
  nofootinbib,
  floatfix
]{revtex4-2}

\usepackage{CJK}
\usepackage[T1]{fontenc}
\usepackage{amsmath,amssymb,bm}
\usepackage{graphicx}
\usepackage{subcaption}
\usepackage{stackengine}
\usepackage{booktabs}
\usepackage{xcolor,braket}
\usepackage{hyperref}
\usepackage{ragged2e}

\hypersetup{
  colorlinks=true,
  linkcolor=blue,
  citecolor=blue,
  urlcolor=blue
}

\graphicspath{{./}}

\newcommand{\insertfigure}[2][0.98\linewidth]{%
  \IfFileExists{#2}{\includegraphics[width=#1]{#2}}{%
    \fbox{\parbox[c][0.22\textheight][c]{0.94\linewidth}{\centering
      Figure placeholder\\[2pt]\texttt{#2}}}%
  }%
}

\begin{document}
\begin{CJK}{UTF8}{gbsn}

\preprint{APS/123-QED}

\title{Quantum Transitions From Integrable to Chaotic}

\author{Sizhe Yan (闫思哲)}
\affiliation{International Center for Quantum Materials, School of Physics, Peking University, Beijing 100871, China}

\author{Biao Wu (吴飙)}
\email{wubiao@pku.edu.cn}
\affiliation{International Center for Quantum Materials, School of Physics, Peking University, Beijing 100871, China}
\affiliation{Wilczek Quantum Center, Shanghai Institute for Advanced Studies, 
University of Science and Technology of China, Shanghai 201315, China}
\affiliation{Hefei National Laboratory, Hefei 230088, China}

\date{\today}

\begin{abstract}
The quantum integrable-to-chaotic transition is commonly pictured as the quantum counterpart of the classical Kolmogorov-Arnold-Moser (KAM) scenario: 
a phaseout transition in which a family of regular eigenstates coexists with, and gradually yields to, chaotic eigenstates. Here we show that 
this is not the only fundamental route from integrability to chaos in quantum systems. We identify a second elementary transition---the morphing transition---which has 
no classical counterpart: all eigenstates hybridize and turn chaotic concurrently, so that no regular subsequence survives at any stage of the crossover. 
The two transitions are spectrally distinct: the phaseout spectrum is an independent superposition of regular and chaotic subsequences with Berry-Robnik statistics 
and a finite short-spacing intercept $P(0)$, whereas the morphing spectrum exhibits global level repulsion with Lenz-Haake statistics and $P(0)=0$. We demonstrate 
both transitions, and their mixture, in three models. The results for the deformed boson model further show that 
the quantum phaseout scenario is conditional on a smooth  quantum-classical correspondence.
\end{abstract}

\maketitle
\end{CJK}

\renewcommand{\figurename}{FIG.}
\renewcommand{\tablename}{TABLE}

\makeatletter
\long\def\@makecaption#1#2{%
  \par
  \vskip\abovecaptionskip
  {\small
   \justifying
   \noindent #1.\enspace #2\par}
  \vskip\belowcaptionskip
}
\makeatother

\section{Introduction}

The Kolmogorov-Arnold-Moser (KAM) theorem describes how classical integrability breaks down: 
for a near-integrable Hamiltonian $H=H_0+\epsilon V$, a positive-measure set of invariant tori survives a sufficiently small 
but finite perturbation, and as $\epsilon$ grows the tori break up successively while the chaotic sea expands~\cite{Poeschel2001,Chirikov1979,LichtenbergLieberman1992}. 
Upon quantization, the surviving tori support a family of regular eigenstates that coexists with, 
and gradually yields to, chaotic eigenstates~\cite{Percival1973,Berry1977,BerryRobnik1984,Yin2019PRE,WangQian2023PRE,WangQian2024PRE}. 
This quantum counterpart of the classical KAM scenario---a transition in which the regular eigenstate 
family is gradually \emph{phased out}---has long 
been the prevailing picture of the quantum integrable-to-chaotic transition.

In this work, we show that the phaseout transition is not the only fundamental route from integrability 
to chaos in quantum systems. We identify a second elementary transition, which we call 
the \emph{morphing} transition: as integrability is broken, all eigenstates hybridize and 
turn chaotic concurrently, so that no regular eigenstate family survives at any stage of the crossover. 
This transition has no classical counterpart and lies beyond the reach of KAM theory. 
The two transitions are spectrally distinct: in the phaseout transition the spectrum is an approximately independent 
superposition of regular and chaotic subsequences described by the Berry-Robnik distribution 
with a finite short-spacing intercept $P(0)$~\cite{BerryRobnik1984}, whereas in the 
morphing transition level repulsion extends across the whole spectrum, 
the statistics follow the Lenz-Haake distribution, and $P(0)$ vanishes~\cite{LenzHaake1991}. 
A generic quantum system may exhibit a mixture of the two.

We demonstrate both transitions, and their mixture, in three models. A three-mode boson model undergoes a phaseout 
transition: its regular eigenstates are localized on surviving classical islands and accurately reproduced 
by a quantum normal form, and the regular subsequence persists~\cite{NakerstHaque2023,CastroEtAl2024}. 
The perturbed XXZ spin-1/2 chain undergoes a morphing transition with no surviving regular subsequence. 
Finally, a deformed three-mode boson model---in which a lattice-scale perturbation breaks 
the smooth quantum-classical correspondence while leaving the classical limit unchanged---realizes a mixture: 
the regular subsequence erodes even when the global degree of chaos is held fixed. 
These results establish the morphing transition as a second elementary paradigm of quantum 
integrability breaking, and show that the phaseout (quantum KAM) scenario is conditional 
on a smooth  quantum-classical correspondence.

\section{Two Elementary Integrable-to-Chaotic Transitions}
\label{theo}
We consider a typical KAM Hamiltonian 
\begin{equation}
H=H_0+\epsilon V\,,
\end{equation}
where $H_0$ is an integrable system with more than one degree of freedom and $V$ is a perturbation of strength $\epsilon$. 
Classically, the KAM theorem guarantees the survival of a positive-measure 
set of nonresonant invariant tori for a sufficiently small but finite $\epsilon$~\cite{Poeschel2001,Chirikov1979,LichtenbergLieberman1992}.
Poincar\'e sections show that the phase space is then divided into islands of invariant tori and chaotic seas; 
as the perturbation strength $\epsilon$ increases, the invariant tori gradually break up, 
the chaotic seas expand, and the system eventually becomes fully chaotic. 

A quantum system described by $H$ also undergoes an integrable-to-chaotic transition, but the transition 
can proceed in different ways. One elementary type is similar to the KAM transition in classical dynamics, 
and we call it the phaseout transition~\cite{BerryRobnik1984,Yin2019PRE}. 
The other elementary type has no classical counterpart, and we call it the morphing transition. 

In general, a quantum system consists of many subsystems. Let $D_{\rm sub}$ be the 
Hilbert-space dimension of each subsystem 
and $L$ the number of subsystems. The phaseout transition is expected 
in systems with $D_{\rm sub}\gg 1$ and $L\sim 1$, while
the morphing transition is expected in systems with $D_{\rm sub}\sim 1$ and $L\gg 1$. 

These two scenarios need not exhaust all possible integrable-to-chaotic transitions. In a concrete system,
the actual transition may instead be a combination of the two. For example, the spectrum may 
consist of an independent superposition of integrable levels and globally hybridized levels; 
as the degree of chaos increases, the fraction of integrable levels decreases, 
while the hybridization among the globally hybridized levels is simultaneously enhanced.
Three examples illustrating the phaseout 
transition, the morphing transition, and a mixture of them are presented in the subsequent sections.

In the following discussion, we do not consider atypical eigenstates such as quantum 
scar states~\cite{SerbynAbaninPapic2021,SchecterIadecola2019,TurnerEtAl2018PXP}, whose measure vanishes: 
the ratio of their number to the total Hilbert-space 
dimension $D$ tends to zero as $D\rightarrow \infty$.

To illustrate the two distinct integrable-to-chaotic transitions, 
we employ two idealized symbolic constructions. They are not intended as concrete microscopic 
models for the systems studied later; their purpose is
to distinguish two organizations 
of the integrable-to-chaotic transition: a phaseout regime in which regular and chaotic levels form approximately 
independent subsequences, and a morphing regime in which level mixing spreads throughout the bulk spectrum.

\subsection{Phaseout Transition}
In the phaseout transition, the Hamiltonian $H$ of the quantum system can be symbolically 
expressed as a block matrix 
\begin{equation}
H_{\rm po}(\rho)
=
\begin{pmatrix}
H_{\mathrm{regular}}^{(1-\rho)} & 0\\
0 & H_{\mathrm{chaotic}}^{(\rho)}
\end{pmatrix},
\end{equation}
where $\rho\in[0,1]$ quantifies the crossover from a fully integrable level structure to a fully chaotic one. 
Specifically, $\rho$ is roughly the ratio of the size of the block $H_{\mathrm{chaotic}}^{(\rho)}$ 
to the size of the full matrix $H_{\rm po}$. When $\rho=0$, the block $H_{\mathrm{chaotic}}$ is absent; when $\rho=1$, the block $H_{\mathrm{regular}}$ is absent. 

The eigenvalues $\mathcal{E}_{\mathrm{po}}(\rho)$ of  $H_{\rm po}$ evidently consist of two statistically independent subsequences, 
\begin{equation}
    \mathcal{E}_{\mathrm{po}}(\rho)
    =
    \mathcal{E}_{\mathrm{regular}}^{(1-\rho)}
    \cup
    \mathcal{E}_{\mathrm{chaotic}}^{(\rho)},
\end{equation}
where the level spacings of $\mathcal{E}_{\mathrm{regular}}^{(1-\rho)}$  obey Poisson statistics 
and those of $\mathcal{E}_{\mathrm{chaotic}}^{(\rho)}$ follow Wigner-Dyson statistics.  
The level spacings of the full spectrum are then well described 
by a Berry-Robnik (BR) distribution~\cite{BerryRobnik1984,ProsenRobnik1994,Prosen1998}.
\begin{equation}
\begin{aligned}
P_{\mathrm{BR}}(s;\rho)
&=e^{-(1-\rho)s}
\Bigg[
(1-\rho)^2
\operatorname{erfc}
\left(
\frac{\sqrt{\pi}}{2}\rho s
\right)
\\
&+
\left(
2\rho(1-\rho)+\frac{\pi}{2}\rho^3s
\right)
\exp{(-\pi \rho^2s^2/4)}
\Bigg].
\end{aligned}
\end{equation}
This follows from the independent superposition of a regular and a chaotic subspectrum. 

\subsection{Morphing Transition}

In the morphing transition, the total Hamiltonian can be viewed as a continuous interpolation 
between an integrable Hamiltonian and a chaotic Hamiltonian. Symbolically, this 
is expressed as 
\begin{equation}
    H_{\mathrm{mo}}(\rho)
    =
    (1-\rho)H_{\mathrm{regular}}
    +\rho H_{\mathrm{chaotic}},
\end{equation}
where the energy levels of $H_{\mathrm{regular}}$ obey Poisson statistics and
$H_{\mathrm{chaotic}}$ is a normalized real symmetric random matrix with Wigner-Dyson statistics. 
All three matrices $H_{\mathrm{mo}}(\rho)$, $H_{\mathrm{regular}}$, 
and $H_{\mathrm{chaotic}}$ have the same dimension. In this transition, at any given value of $\rho$, 
there is only a single set of eigenstates $\mathcal{E}_{\mathrm{morph}}(\rho)$, whose level spacings roughly 
obey the Lenz-Haake distribution~\cite{LenzHaake1991}
\begin{equation}
P_{\mathrm{LH}}(s;\eta)
=\mu(\eta)\,p_R\!\left(\mu(\eta)s;\eta\right),
\end{equation}
where 
\begin{align}
p_R(r;\eta)
&=
\int_0^\infty e^{-x}
\frac{r}{2\eta^2}
\exp\!\left[-\frac{r^2+x^2}{4\eta^2}\right]
I_0\!\left(\frac{rx}{2\eta^2}\right)\,dx,
\\
\mu(\eta)&=\int_0^\infty r\,p_R(r;\eta)\,dr \,.
\end{align} 
Here \(I_0\) is the modified Bessel function of the first kind of order zero. The limiting 
values \(\eta=0\) and \(\eta=\infty\) correspond to the integrable and fully chaotic regimes, respectively.
The Lenz-Haake (LH) distribution is obtained by considering two neighboring Poisson levels 
coupled by a real symmetric random perturbation~\cite{LenzHaake1991}. 
It is exact for a two-level system. For a multilevel system, however, it does not account for the influence of couplings to additional surrounding levels, and may therefore be regarded here 
as a low-order approximation. 

\begin{figure}[htbp]
  \centering
  \begin{subfigure}{0.49\columnwidth}
    \centering
    \phantomcaption
    \stackinset{l}{1pt}{t}{1pt}{\footnotesize(\thesubfigure)}{%
      \includegraphics[width=\linewidth]{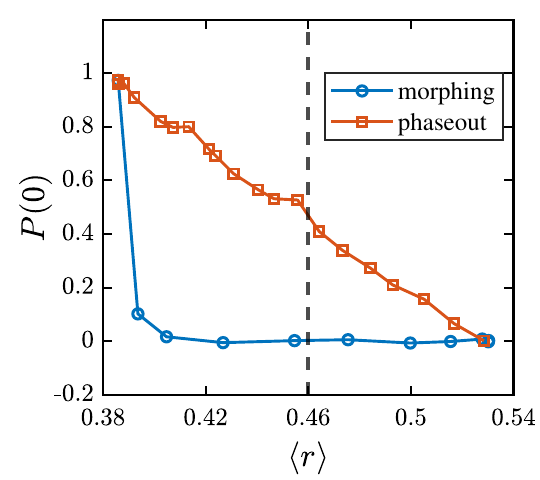}}
    \label{fig:RPa}
  \end{subfigure}
  \hfill
  \begin{subfigure}{0.49\columnwidth}
    \centering
    \phantomcaption
    \stackinset{l}{1pt}{t}{1pt}{\footnotesize(\thesubfigure)}{%
      \includegraphics[width=\linewidth]{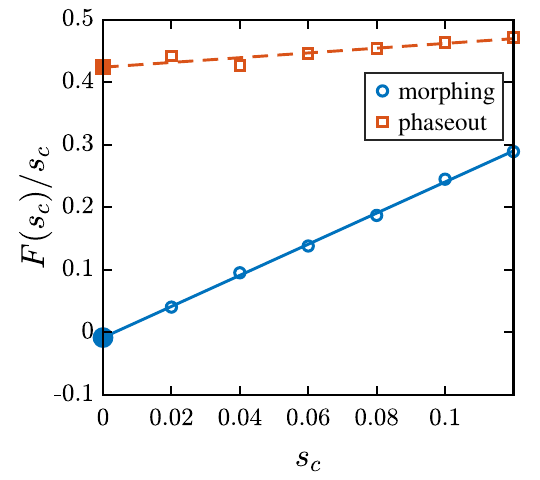}}
    \label{fig:RPb}
  \end{subfigure}
  \vspace{0.3em}
  \begin{subfigure}{0.49\columnwidth}
    \centering
    \phantomcaption
    \stackinset{l}{1pt}{t}{1pt}{\footnotesize(\thesubfigure)}{%
      \includegraphics[width=\linewidth]{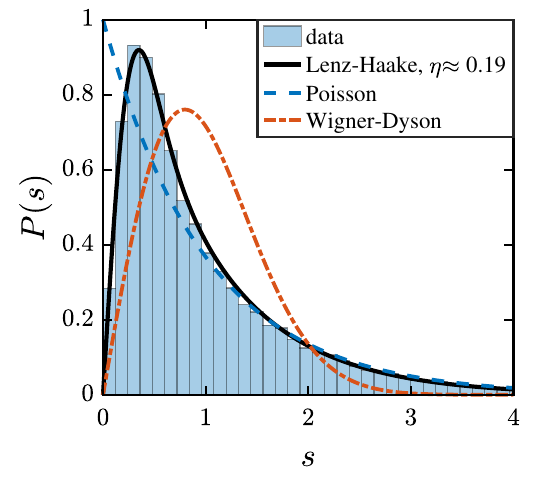}}
    \label{fig:RPc}
  \end{subfigure}
  \hfill
  \begin{subfigure}{0.49\columnwidth}
    \centering
    \phantomcaption
    \stackinset{l}{1pt}{t}{1pt}{\footnotesize(\thesubfigure)}{%
      \includegraphics[width=\linewidth]{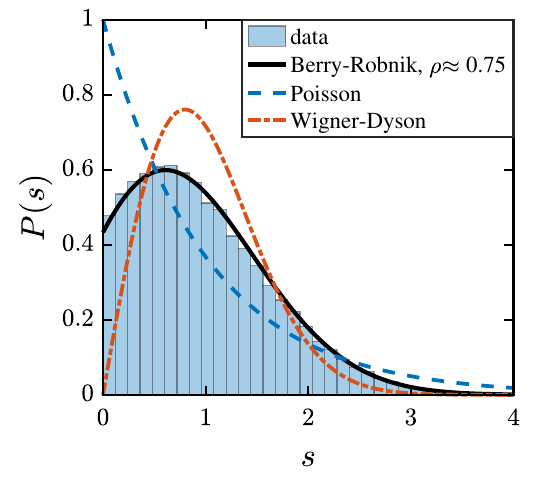}}
    \label{fig:RPd}
  \end{subfigure}
  \caption{(a) Variation of \(P(0)\) with \(\langle r\rangle\) in two different quantum systems, which undergo 
  distinct integrable-to-chaotic transitions. (b) Schematic of the linear fit of \(P(0)\) at \(\langle r\rangle_*=0.46\) for the two Hamiltonians. (c) Level-spacing statistics of \(H_{\mathrm{mo}}\), corresponding to a Lenz-Haake fit. (d) Level-spacing statistics of \(H_{\mathrm{po}}\), corresponding to a Berry-Robnik fit.}
  \label{fig:RP}
\end{figure}

The most unambiguous distinction between the BR and LH distributions is the value of the intercept, 
namely \(P(0)\). For the BR distribution, \(P(0)=1-\rho^2\) varies gradually, whereas 
for the LH distribution, as soon as the system leaves the integrable regime, \(P(0)\) 
drops immediately to zero and thereafter remains unchanged, as shown in Fig. \ref{fig:RP}. 
The underlying reason is that, in the phaseout transition, 
there exist nearly independent integrable levels that exhibit no level repulsion either among themselves 
or with the chaotic levels, thereby retaining a finite \(P(0)\); in the morphing transition, by contrast, 
repulsion is present among all levels, and even if it is initially weak, it suffices to eliminate \(P(0)\) entirely. 

\subsection{Quantum Normal Forms}
When the perturbation strength $\epsilon$ is sufficiently small, the KAM theorem 
guarantees the survival of certain invariant tori. Classically, this is seen by 
eliminating the nonresonant harmonics of \(V\) order by order via canonical transformations 
obeying the homological equation~\cite{Deprit1969,Hori1966}: with the small divisors tamed by Diophantine conditions, 
the iteration converges on a Cantor set of surviving invariant tori~\cite{Poschel1982,Poeschel2001,Arnold1989}. Near a resonance, 
the resonant harmonics cannot be removed and are retained in an integrable resonant normal form 
describing the associated island~\cite{SandersVerhulstMurdock2007,LichtenbergLieberman1992}, 
whose persistence in the full system must nevertheless be assessed independently.

There are similar results for quantum systems undergoing the phaseout transition. Without loss of 
generality, we consider a quantum system with two degrees of freedom, 
described by
\begin{align}
    &\hat H=\hat H_0(\hat{\bm I})+\varepsilon\hat V(\hat{\bm I},\hat{\bm \theta})
    \;\\
    &\hat H_0\lvert \bm n\rangle=E(\bm n)\lvert \bm n\rangle,
    \;
    {\bm n}=(n_1,n_2).
\end{align}
The associated actions are \({I}_{i}=\hbar({ n_i}+\nu)\)~\cite{Keller1958}. 

The operator \(\hat{\bm \theta}\) is not well defined, 
but the action shift operators are. Introduce the elementary shift operators
\begin{equation}
    \hat T_1\lvert n_1,n_2\rangle
    =
    \lvert n_1+1,n_2\rangle,
    \;
    \hat T_2\lvert n_1,n_2\rangle
    =
    \lvert n_1,n_2+1\rangle.
\end{equation}
Negative powers denote the corresponding adjoint shifts. For a displacement
\({\bm m}=(m_1,m_2)\), write
\(
    \hat {\bm T}^{\bm m}\equiv\hat T_1^{m_1}\hat T_2^{m_2}.
\)
The normal-form construction is local to the interior of the allowed
quantum-number lattice, where the shift paths are well defined; boundary
terms can be treated separately. Since \(\hat{\bm T}\) acts as a translation operator in action space,
\(
    \hat T_j=e^{i\hat\theta_j},
\)
we have the commutation relations
\begin{equation}
    [\hat n_j,\hat {\bm T}^{\bm m}]
    =
    m_j\hat {\bm T}^{\bm m}.
\end{equation}
Furthermore, the perturbation operator can be expanded in terms of the shift operators.
\begin{equation}
    \hat V
    =
    \sum_{{\bm n},{\bm m}}v_{\bm m}({\bm n})\hat D_{\bm n}\hat {\bm T}^{\bm m},
    \;
    v_{\bm m}({\bm n})=\langle {\bm n}\vert\hat V\vert {\bm n}-{\bm m}\rangle,
\end{equation}
with \(\hat D_{\bm n}=\lvert {\bm n}\rangle\langle {\bm n}\rvert\). This corresponds to the Fourier expansion of the classical case.

A quantum normal form is obtained by a unitary conjugation~\cite{Anikin2009,BravyiDiVincenzoLoss2011}
\begin{equation}
    \hat H'
    =
    e^{-i\hat W}\hat H e^{i\hat W},
    \qquad
    \hat W=\hat W^\dagger.
\end{equation}
To first order, selected nonresonant matrix elements are eliminated by imposing
\begin{equation}
    -i[\hat W,\hat H_0]
    +
    \varepsilon\hat V
    =
    0.
\end{equation}
Writing
\begin{equation}
    \hat W
    =
    \sum_{{\bm n},{\bm m}}w_{\bm m}({\bm n})\hat D_{\bm n}\hat {\bm T}^{{\bm m}},
\end{equation}
one obtains
\begin{equation}
    w_{\bm m}({\bm n})
    =
    i\,\frac{\varepsilon v_{\bm m}({\bm n})}
    {E({\bm n})-E({\bm n}-{\bm m})}.
\end{equation}
This is the quantum counterpart of the classical homological equation: the
Fourier harmonic is replaced by a lattice shift, and its frequency divisor
is replaced by the corresponding energy difference. In the semiclassical limit, \(E({\bm n})\) becomes
a smooth function, and the energy differences reduce
to the classical frequencies \({\bm m}\cdot\bm\omega({\bm I})\). In this way, 
one recovers the classical normal-form procedure~\cite{Popov2000,VuNgoc2009}.

For simplicity, we consider the \(1\!:\!-1\) resonance. In this case, the retained quantum transitions are
\(
    \hat {\bm T}^{(m,m)}
    =
    \hat T_1^m\hat T_2^m,
\)
which preserve
\(
    \hat K=\hat n_1-\hat n_2.
\)
One may define \(
|K,J\rangle
=
\left|
n_1=\frac{J+K}{2},
n_2=\frac{J-K}{2}
\right\rangle
\), and then write
\begin{equation}
    \begin{aligned}
        \hat H=&\sum_{K}\sum_{J}E\left(\frac{J+K}{2},\frac{J-K}{2}\right)\lvert K,J\rangle\langle K,J\rvert\\
+&\varepsilon
\sum_K\sum_J\sum_{r\neq0}
t_r(K,J)
\lvert K,J+2r\rangle\langle K,J\rvert\\
+&\hat H_{\mathrm{nr}}
    \end{aligned}
\end{equation}
where \(\hat H_{\mathrm{nr}}\) denotes the remaining nonresonant perturbation and
\begin{equation}
    t_r(K,J)
    =
    v_{(r,r)}
    \left(
        \frac{J+K}{2}+r,\,
        \frac{J-K}{2}+r
    \right).
\end{equation}
Consistent with the classical normal form, the first two terms also constitute an integrable Hamiltonian~\cite{Anikin2009,Popov2000}. The remaining non-resonant terms can be successively 
eliminated via an iterative procedure, thereby yielding an approximately diagonal Hamiltonian. 
The normal-form analysis thus reveals a remarkable formal consistency between the classical and quantum descriptions, and the mixed integrable-chaotic structure of the classical phase space is reflected 
in the distinct types of eigenstates of the quantum Hamiltonian~\cite{Percival1973,Berry1977,BerryRobnik1984,NakerstHaque2023,CastroEtAl2024}. 
The quantization of KAM theory and of the normal-form language depends crucially on the 
quantum-classical correspondence of the system: only when such a correspondence exists 
can a positive-measure family of regular eigenstates be approximated, order by order, 
by a quantum normal form.

\section{Numerical Techniques}
\label{sec:num}
In the following sections, we present numerical results for concrete models 
that demonstrate the above theoretical discussion. Two techniques are used for all the models: 
a method for estimating the intercept $P(0)$ of the level-spacing distribution, and a common variable 
that measures the degree of chaos in different quantum systems. 

\noindent {\it Computation of $P(0)$ --} As mentioned above, 
The most prominent distinction between the BR and LH distributions is the value of the intercept, 
namely \(P(0)\). 
Therefore, computing \(P(0)\) provides an effective means of distinguishing the two integrable-to-chaotic 
transitions. However, the computation is complicated by that the continuous level-spacing distribution 
in fact describes an infinitely dense spectrum. For a finite system, 
the empirical \(P(s)\) is discrete, and one can only fit an approximate function. 
As a result, \(P(0)\) 
cannot be computed directly, but must instead be estimated by extrapolation. 

For a spectrum with $N_{\rm t}$ level spacings, we estimate the intercept from 
the cumulative short-spacing weight, which is defined as 
\begin{equation}
    F(s_c)
    =
    \frac{\#\{\mathrm{level~spacings}<s_c\}}{N_{\mathrm{t}}}\,,
\end{equation}
where \(s_c\) is chosen to be small relative to the typical level spacings in the spectrum. 
\(P(0)\), as the intercept of a continuous function, is defined in the successive limits as
\begin{equation}
P(0)
=
\lim_{s_c\to0^+}
\lim_{N_{\mathrm{t}}\to\infty}
\frac{ F(s_c)}{s_c}.
\end{equation}
In computation, we estimate \(P(0)\) with the following linear extrapolation
\begin{equation}
    \frac{F(s_c)}{s_c}
    =P(0)+O(s_c)\,.
\end{equation}

\noindent {\it Adjacent-gap ratio \(\langle r\rangle\)--} To study the integrable-to-chaotic transition, 
one starts from an integrable model and introduces a 
perturbation that progressively destroys its integrability until the system becomes fully chaotic. Particular attention 
is paid to the intermediate regime in which the perturbation is neither too weak nor too strong, 
so that the system is neither fully integrable nor fully chaotic. 

To quantify the degree of chaos in the system, 
we adopt the average adjacent-gap ratio \(\langle r\rangle\)~\cite{OganesyanHuse2007,AtasEtAl2013} as an indicator. For a fully integrable system, 
the level-spacing distribution is Poissonian, with the corresponding value 
\(\langle r\rangle_{\mathrm{integrable}}=2\ln 2-1\approx 0.3863\); for a fully chaotic system (GOE), the level-spacing distribution is of 
Wigner-Dyson type, with \(\langle r\rangle_{\mathrm{chaotic}}\approx 0.5307\). When $\langle r\rangle$ 
approaches the midpoint of \(\langle r\rangle_{\mathrm{integrable}}\) 
and \(\langle r\rangle_{\mathrm{chaotic}}\), we regard the system as having reached the desired intermediate 
regime. In this work, without loss of generality, we take the target value to be \(\langle r\rangle_*=0.46\).

\section{Three-Mode Boson Model}
We use a three-mode boson model to demonstrate the phaseout transition. 
It is representative of quantum systems that have a well-defined classical counterpart. 

\subsection{Model and Semiclassical Limit}
The three modes in this boson system are labelled by \(\mu=0,+,-\), and its 
Hamiltonian is~\cite{EvrardEtAl2024SpinorScars}
\begin{equation}
\hat H_{b}(p)=\hat H_0+p\hat V,
\label{eq:bsc-hamiltonian}
\end{equation}
where
\begin{equation}
\hat H_0=
\frac{c}{N}
\left[
\hat n_0(N-\hat n_0)
+\gamma(\hat n_+-\hat n_-)^2
\right]
+\alpha\hat n_0,
\label{eq:bsc-h0}
\end{equation}
and
\begin{equation}
\hat V=
\frac{1}{\sqrt{2}}
\left(
\hat a_+^\dagger\hat a_0
+\hat a_-^\dagger\hat a_0
+\mathrm{H.c.}
\right).
\label{eq:bsc-v}
\end{equation}
Here 
$\hat n_\mu=\hat a_\mu^\dagger\hat a_\mu$ is the number operator for mode $\mu$. The 
total number of bosons is conserved, i.e., $\hat n_0+\hat n_++\hat n_-=N$, where $N$ is
the total particle number. The Hilbert-space dimension of the system is
\(
D=(N+1)(N+2)/2.
\)

At \(p=0\), the Hamiltonian is diagonal in the Fock basis. Defining the
imbalance operator
\(
\hat m_q=\hat n_+-\hat n_-,
\) 
the eigenstates are labelled by \(N_0\), the number of bosons in mode $0$,  and \(M_q\), the particle-number difference
between the other two modes. The corresponding eigenenergies are
\begin{equation}
E^{(0)}({N_0,M_q})
=
\frac{c}{N}
\left[
N_0(N-N_0)+\gamma M_q^2
\right]
+\alpha N_0.
\label{eq:bsc-integrable-spectrum}
\end{equation}
Clearly, \(N_0\) and \(M_q\) are conserved when \(p=0\). 
For \(p\neq0\),  mode \(0\) is coupled to the other two modes.
The coupling changes \(M_q\), while the \(+\leftrightarrow-\)
exchange parity remains conserved and thus remains a good quantum number.

This system has a classical limit at large $N$, where 
it can be described by the number-conserving coherent states
\begin{equation}
|\bm z;N\rangle=
\frac{1}{\sqrt{N!}}
\left(
z_0\hat a_0^\dagger
+z_+\hat a_+^\dagger
+z_-\hat a_-^\dagger
\right)^N
|\mathrm{vac}\rangle,
\label{eq:bsc-coherent-state}
\end{equation}
with
\(
\sum_{\mu=0,+,-}|z_\mu|^2=1.
\) 
Writing \(z_\mu=\sqrt{n_\mu}e^{-i\phi_\mu}\), we introduce
$m=n_+-n_-$ and the relative phases
\begin{equation}
\vartheta_1=
\phi_0-\frac{\phi_++\phi_-}{2},
\qquad
\vartheta_2=
\frac{\phi_+-\phi_-}{2}.
\end{equation}
The  classical Hamiltonian per particle is
\begin{equation}
\begin{aligned}
&h(n_0,m,\vartheta_1,\vartheta_2)\\
=&h_0+
p\sqrt{n_0(1-n_0+m)}\cos(\vartheta_1-\vartheta_2)\\
&+
p\sqrt{n_0(1-n_0-m)}\cos(\vartheta_1+\vartheta_2)\,,
\end{aligned}
\label{eq:classical-bsc}
\end{equation}
where $h_0=c\left[n_0(1-n_0)+\gamma m^2\right]+\alpha n_0$. 

\subsection{Entanglement and Poincar\'e Section}
\begin{figure}[htbp]
  \centering
  \begin{subfigure}{0.49\columnwidth}
    \centering
    \phantomcaption
    \stackinset{l}{1pt}{t}{1pt}{\footnotesize(\thesubfigure)}{%
      \includegraphics[width=\linewidth]{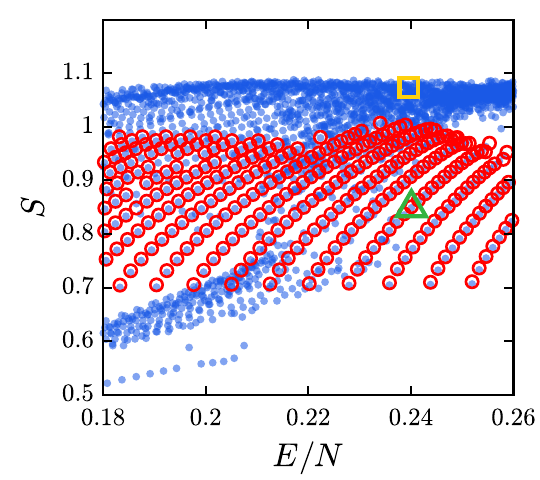}}
    \label{fig:BSCa}
  \end{subfigure}
  \hfill
  \begin{subfigure}{0.49\columnwidth}
    \centering
    \phantomcaption
    \stackinset{l}{1pt}{t}{1pt}{\footnotesize(\thesubfigure)}{%
      \includegraphics[width=\linewidth]{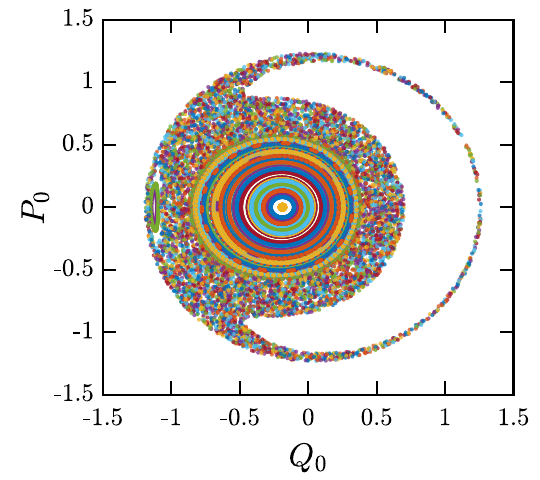}}
    \label{fig:BSCb}
  \end{subfigure}

  \begin{subfigure}{0.49\columnwidth}
    \centering
    \phantomcaption
    \stackinset{l}{1pt}{t}{1pt}{\footnotesize(\thesubfigure)}{%
      \includegraphics[width=\linewidth]{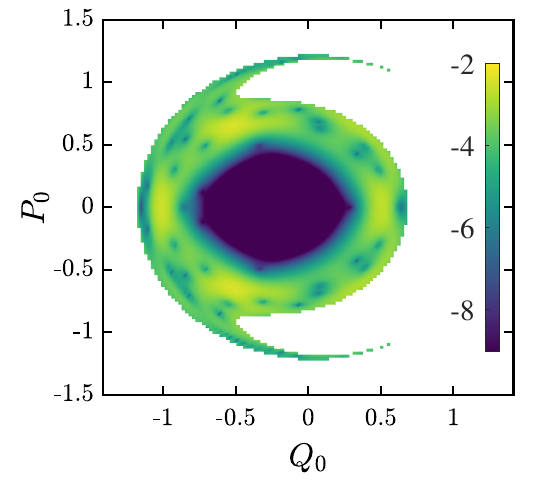}}
    \label{fig:BSCc}
  \end{subfigure}
  \hfill
  \begin{subfigure}{0.49\columnwidth}
    \centering
    \phantomcaption
    \stackinset{l}{1pt}{t}{1pt}{\footnotesize(\thesubfigure)}{%
      \includegraphics[width=\linewidth]{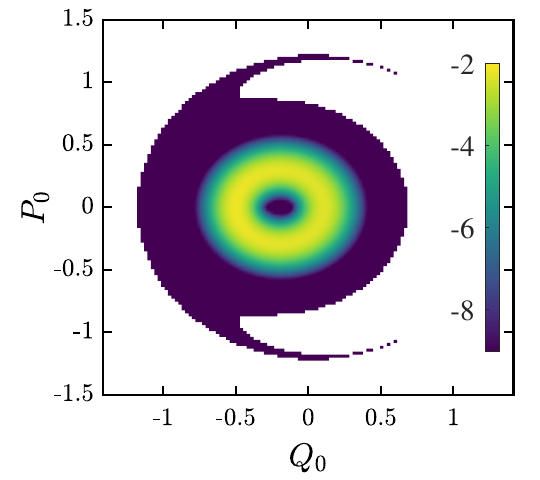}}
    \label{fig:BSCd}
  \end{subfigure}
  \caption{(a) Scatter plot of energy versus entanglement entropy for \(N=180\) and \(p=0.055\). Red circles denote eigenstates obtained from the quantum normal-form approximation. The yellow square and green triangle mark two eigenstates near \(E/N\simeq 0.24\), with level indices \(3193\) and \(3177\), respectively. (b) Poincar\'e surface of section of the classical system at \(p=0.055\), \(E=0.24\), \(\vartheta_2=0\), and \(\dot{\vartheta}_2 >0\). Here \(Q_0=\sqrt{2n_0}\cos\vartheta_1\) and \(P_0=\sqrt{2n_0}\sin\vartheta_1\). A prominent integrable island is visible in the figure, corresponding to a \(1:1\) or \(1:-1\) resonance. (c) Husimi projection of the \(3193\)th eigenstate onto the Poincar\'e section, corresponding to the yellow square in panel (a). (d) Husimi projection of the \(3177\)th eigenstate onto the Poincar\'e section, corresponding to the green triangle in panel (a).}
  \label{fig:BSC}
\end{figure}

Our numerical results show that this boson system undergoes a phaseout transition as $p$ increases from zero. 
As stated in Sec.~\ref{sec:num}, we present our numerical results at the crossover point \(\langle r\rangle_*=0.46\), 
which for this system occurs at the perturbation strength \(p=0.055\). 
We have computed the entanglement entropy for all the eigenstates in the even-parity sector with 
\(
c=1,\;
\alpha=0.03,\;
\gamma=\frac{\sqrt{5}}{4}
\), and the results are plotted in Fig.~\ref{fig:BSCa}. The scatter plot 
consists of an upper chaotic cloud together with a Peres-lattice structure~\cite{Peres1984} below, a clear
indication that there are two subsequences of eigenstates: one regular and the other chaotic. 
The corresponding classical model exhibits the same coexistence of regular and chaotic motion:  
as shown in Fig.~\ref{fig:BSCb}, its Poincar\'e section 
contains both chaotic regions and island-like integrable regions. 

We have selected two eigenstates, the 3193rd and the 3177th. The 3193rd eigenstate belongs to the 
chaotic subsequence and is marked with a yellow square in Fig.~\ref{fig:BSCa}; 
the 3177th eigenstate belongs to the 
regular subsequence and is marked with a green triangle. 
Figures~\ref{fig:BSCc} and \ref{fig:BSCd} show the Husimi projections of these two eigenstates 
onto the Poincar\'e surface of section. To construct the Husimi projection on the Poincar\'e section, we evaluate the coherent-state overlap \(Q_{\mathrm{sec}}=|\langle\bm z;N|\psi\rangle|^2\) at \(\vartheta_2=0\), with \(m=m(E,Q_0,P_0,\vartheta_2=0)\) determined by the classical energy constraint on the branch satisfying \(\dot{\vartheta}_2>0\). The resulting intensity is plotted in the \((Q_0,P_0)\) plane on a logarithmic color scale, \(\log_{10}Q_{\mathrm{sec}}\). The regular 3177th eigenstate is localized on the surviving invariant tori 
within the regular island, whereas the chaotic 3193rd eigenstate spreads throughout the chaotic sea. 

\subsection{Spectral Statistics}
\begin{figure}[htbp]
  \centering
  \begin{subfigure}{0.46\columnwidth}
    \centering
    \phantomcaption
    \stackinset{l}{-8pt}{t}{1pt}{\footnotesize(\thesubfigure)}{%
      \includegraphics[width=\linewidth]{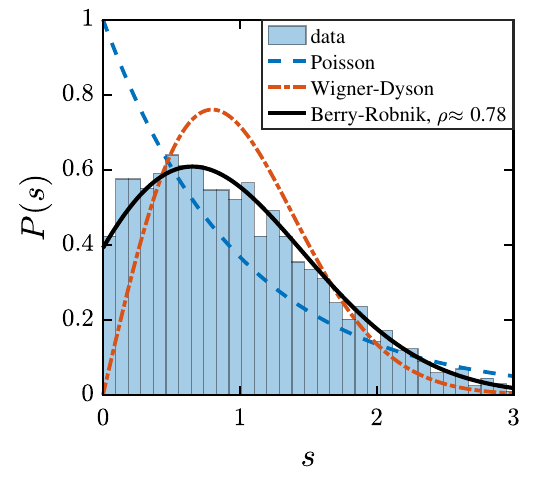}}
    \label{fig:BSC2a}
  \end{subfigure}
  \hfill
  \begin{subfigure}{0.46\columnwidth}
    \centering
    \phantomcaption
    \stackinset{l}{-8pt}{t}{1pt}{\footnotesize(\thesubfigure)}{%
      \includegraphics[width=\linewidth]{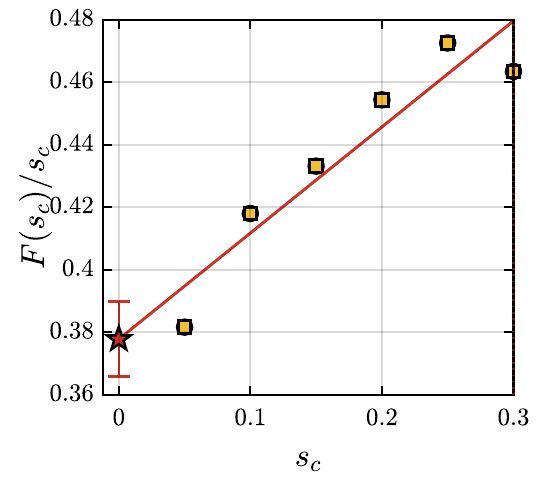}}
    \label{fig:BSC2b}
  \end{subfigure}

  \vspace{0.8em}
  \begin{subfigure}{0.46\columnwidth}
    \centering
    \phantomcaption
    \stackinset{l}{-8pt}{t}{1pt}{\footnotesize(\thesubfigure)}{%
      \includegraphics[width=\linewidth]{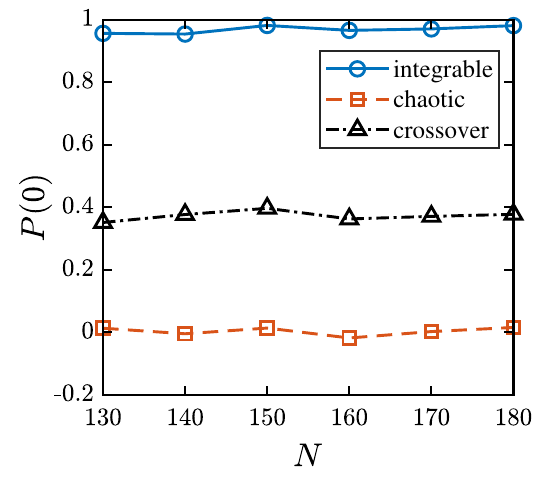}}
    \label{fig:BSC2c}
  \end{subfigure}
  \hfill
  \begin{subfigure}{0.46\columnwidth}
    \centering
    \phantomcaption
    \stackinset{l}{-8pt}{t}{1pt}{\footnotesize(\thesubfigure)}{%
      \includegraphics[width=\linewidth]{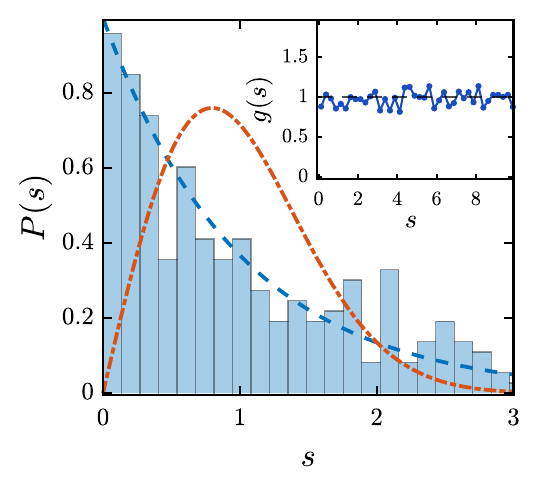}}
    \label{fig:BSC2d}
  \end{subfigure}
  \caption{(a) For \(N=180\) and \(p=0.055\), the level statistics are evaluated in the energy window \(E/N\in[0.18,0.26]\).  (b) An extrapolation fit yields \(P(0)=0.38\pm0.012\). (c) \(P(0)\) of the system for different particle numbers in three cases: \(\langle r \rangle_* = 0.46\), integrable, and fully chaotic. (d) For the integrable levels, their spectral statistics and their cross-spectral correlations with the remaining levels}
  \label{fig:BSC2}
\end{figure}

The phaseout transition from integrability to chaos in a quantum system is characterized by a 
decrease in the fraction of regular eigenstates, accompanied by an increase in the fraction of chaotic ones. 
This is also reflected in the spectral statistics: the energy levels fall into two families, one following 
Poisson statistics and the other following Wigner-Dyson statistics, 
and the combined statistics obey the Berry-Robnik distribution, as 
seen in Fig.~\ref{fig:BSC2a}. 

This feature is most clearly manifested in the intercept \(P(0)\) of the level-spacing distribution. 
When the system lies between integrability and chaos, 
the level-spacing distribution remains finite and strictly positive at $s=0$, as shown in Fig.~\ref{fig:BSC2b}. 
As the particle number $N$ is varied, \(P(0)\) remains essentially 
unchanged, as shown in Fig.~\ref{fig:BSC2c}, indicating that the fraction of regular states is preserved. 
This is also expected from the existence of a classical counterpart of the system.

To further demonstrate that this spectral statistics reflects the persistence of the integrable structure 
guaranteed by KAM theory, we compute the associated regular eigenstates approximately 
by means of a quantum normal form. On the energy-entanglement-entropy scatter plot, these states 
are reproduced to a high degree of accuracy, as shown in Fig.~\ref{fig:BSCa}, 
indicating that the regular eigenstates arise from the quantization of the residual invariant tori of the classical system.

To demonstrate that the energy levels of the regular eigenstates indeed correspond 
to a set of nearly independent levels, we isolate the energies of all identified regular 
eigenstates and examine their statistics. As shown in Fig.~\ref{fig:BSC2d}, the level-spacing distribution 
of these levels is well fitted by the Poisson distribution, indicating that the levels themselves 
form an integrable subspectrum. To test whether the identified regular levels are statistically correlated 
with the remaining spectrum, we compute the normalized cross-spectral correlation
\begin{equation}
\begin{aligned}
        g(s)&=
\frac{\langle\rho_A(x)\rho_B(x+s)\rangle_x}
{\langle\rho_A(x)\rangle_x\,
 \langle\rho_B(x)\rangle_x},
\\
\rho_{A,B}(x)&=\sum_{i\in A,B}\delta(x-x_i),
\end{aligned}
\end{equation}
where the subsequences $A$ and $B$ are obtained by classifying the commonly unfolded spectrum. 
Statistical independence gives \(g(s)=1\). 
As shown in Fig. \ref{fig:BSC2d}, the correlation remains close to unity at small spacings, with no significant depletion, 
supporting approximate independence between the identified regular levels and the remaining spectrum.

The integrable-to-chaotic transition seen in this boson system is therefore spectrally consistent 
with the phaseout transition. The regular states identified from the quantum 
normal form constitute an approximately Poissonian subsequence, while the remaining states display chaotic level statistics. 
The finite value of \(P(0)\), 
together with the absence of a strong depletion in the cross-spectral correlation, indicates that these two subsequences remain approximately independent.

\section{XXZ Spin Chain}
To illustrate the morphing transition from integrability to chaos, we use an XXZ spin chain 
with a next-nearest-neighbor perturbation. As the perturbation becomes 
stronger, our numerical computations reveal a clear morphing transition. 

\subsection{Perturbed XXZ Model}
The well-known periodic spin-\(1/2\) XXZ chain has the Hamiltonian
\begin{equation}
\hat H_{\mathrm{XXZ}}=
\sum_{j=1}^{L}
\left(\hat S_j^x\hat S_{j+1}^x+
\hat S_j^y\hat S_{j+1}^y+
\Delta\hat S_j^z\hat S_{j+1}^z\right),
\end{equation}
where  \(\Delta\) is the anisotropy parameter. 
This spin chain is perturbed by  a translation-invariant next-nearest-neighbor interaction and the Hamiltonian becomes
\begin{equation}\begin{aligned}
    \hat H(\lambda)&=\hat H_{\mathrm{XXZ}}+\lambda\hat V_2,\\
\hat V_2&=\sum_{j=1}^{L}
\left(\hat S_j^x\hat S_{j+2}^x+
\hat S_j^y\hat S_{j+2}^y+
\Delta\hat S_j^z\hat S_{j+2}^z\right).
\end{aligned}
\end{equation}
This model is exactly solvable 
at \(\lambda=0\)~\cite{Franchini2017XXZ} and at \(\lambda=1/2\)~\cite{MajumdarGhosh1969}. 
For other values of $\lambda$, the spin chain is generally nonintegrable.
At \(\lambda=0\), the nearest-neighbor chain is solvable by the Bethe
ansatz and possesses an extensive hierarchy of conserved charges.
A generic finite \(\lambda\) breaks this integrability while preserving
total magnetization, translation, spatial reflection, and global spin
reversal.

\subsection{Spectral Statistics}
In our numerical computations, we choose even \(L\) for convenience. 
This spin chain has several discrete symmetries. To remove degeneracy crossings between distinct exact
symmetry sectors, we focus on the fully resolved sector
\[
S^z_{\mathrm{tot}}=0,\qquad k=0,\qquad R=+1,\qquad C=+1,
\]
where \(R\) and \(C\) denote reflection and spin-reversal parity, and \(k\) labels the translational-momentum sector under consideration. Resolving these symmetries is essential because crossings between different exact symmetry sectors 
can mimic Poisson-like statistics and obscure the level-repulsion mechanism 
within a single irreducible block. All spectral quantities are therefore evaluated only after projection 
onto the same symmetry sector for every \(L\) and \(\lambda\). For each \(L\), we choose the coupling \(\lambda(L)\) 
such that the system satisfies \(\langle r\rangle_*=0.46\). For \(L=22\), the corresponding coupling is \(\lambda_*=0.066\).
In all computations, we choose \(\Delta=0.55\). 

\begin{figure}[htbp]
  \centering
  \begin{subfigure}{0.49\columnwidth}
    \centering
    \phantomcaption
    \stackinset{l}{1pt}{t}{1pt}{\footnotesize(\thesubfigure)}{%
      \includegraphics[width=\linewidth]{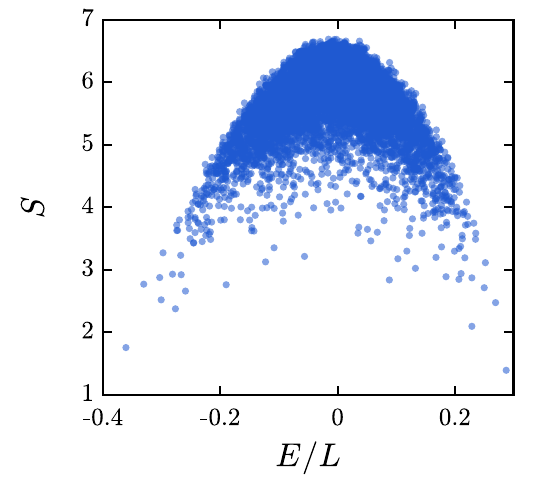}}
    \label{fig:XXZa}
  \end{subfigure}
  \hfill
  \begin{subfigure}{0.49\columnwidth}
    \centering
    \phantomcaption
    \stackinset{l}{1pt}{t}{1pt}{\footnotesize(\thesubfigure)}{%
      \includegraphics[width=\linewidth]{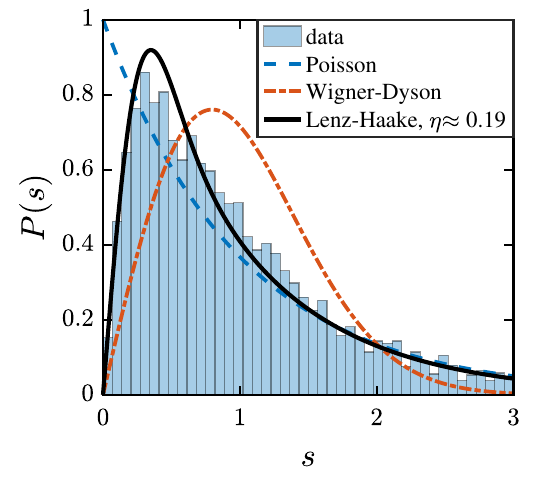}}
    \label{fig:XXZb}
  \end{subfigure}

  \vspace{0.8em}
  \begin{subfigure}{0.49\columnwidth}
    \centering
    \phantomcaption
    \stackinset{l}{1pt}{t}{1pt}{\footnotesize(\thesubfigure)}{%
      \includegraphics[width=\linewidth]{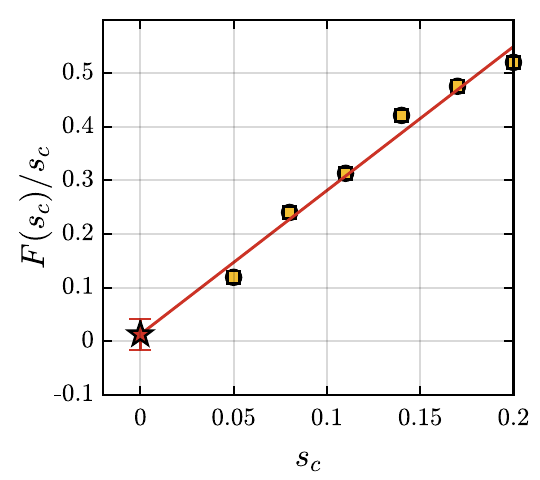}}
    \label{fig:XXZc}
  \end{subfigure}
  \hfill
  \begin{subfigure}{0.49\columnwidth}
    \centering
    \phantomcaption
    \stackinset{l}{1pt}{t}{1pt}{\footnotesize(\thesubfigure)}{%
      \includegraphics[width=\linewidth]{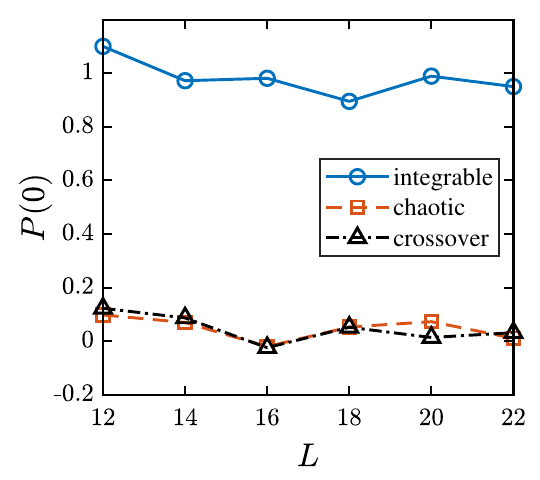}}
    \label{fig:XXZd}
  \end{subfigure}
  \caption{(a) Energy-entanglement entropy scatter plot of the eigenstates at \(L=22\) and \(\lambda=0.066\). 
  One can see that the low-entanglement region does not form a lattice-like structure. 
  (b) Level-spacing statistics in the middle \(80\%\) of the spectrum at this point. (c) By linear extrapolation, 
  we obtain \(P(0)=0.0129\pm 0.0292\). (d) \(P(0)\) of the model at the integrable, chaotic, 
  and crossover points for different particle numbers.}
  \label{fig:XXZ}
\end{figure}

Figure \ref{fig:XXZa} shows the half-chain entanglement entropy as a function of energy for the \(L=22\) spectrum. 
The entropies display a broad spread across the plotted energy range, rather than clustering 
closely around a smooth, nearly thermal envelope. In addition, the eigenstates with 
comparatively low entanglement do not form a clear lattice-like pattern across energy. 
This contrasts with the Peres-lattice-like organization observed in the bosonic trimer, 
where the regular states can be associated with structures in the underlying classical phase space. 
The entanglement plot therefore shows no similarly organized family of low-entanglement states 
in the XXZ crossover spectrum. Entanglement alone, however, does not determine 
whether a regular spectral subsequence exists, so we also examine the level spacings.

Figure \ref{fig:XXZb} shows the unfolded nearest-neighbor spacing statistics in the middle \(80\%\) of the spectrum. 
We estimate the short-spacing intercept \(P(0)\) using the linear extrapolation procedure described above. 
In particular, the estimate for \(L=22\) is \(P(0)=0.0129\pm0.0292\), which is consistent with zero 
within the reported uncertainty; the estimates for the other sizes shown in Fig. \ref{fig:XXZd} are likewise 
consistent with a vanishing intercept. This result contrasts with the nonzero intercept found for the boson model 
over the particle numbers studied.

The vanishing intercept is important because an approximately independent superposition of regular 
Poisson-like levels and chaotic levels can retain a finite density of arbitrarily small spacings, producing \(P(0)>0\). 
By contrast, when level repulsion extends across the spectrum, the spacing distribution is suppressed at the origin 
and \(P(0)\) vanishes. The level-spacing statistics of the XXZ model are well described 
by the Lenz-Haake distribution~\cite{LenzHaake1991}, which also indicates global level repulsion 
during the transition to chaos. This interpretation is consistent with the entanglement data, 
which show no clear lattice-like organization among the low-entanglement states. 
This implies that all eigenstates change concurrently and are converted together into chaotic eigenstates. 

These results show that the XXZ crossover is consistent with the morphing transition. 
The short-spacing intercept is statistically compatible with zero, and the unfolded spacing distribution 
is well described by the Lenz-Haake distribution. These observations indicate that level repulsion extends 
across the resolved spectrum rather than leaving an approximately independent Poisson-like subsequence.

The above observations differ markedly from those for the boson trimer. The vanishing of \(P(0)\) indicates 
that the spectrum does not split into two nearly independent sets of levels, nor does it retain an integrable level structure. 
In other words, during the breakdown of integrability in the XXZ chain, no integrable structure survives 
within a subspace of nonzero measure. This suggests that, in a system without a classical counterpart, 
the breakdown of integrability proceeds through all eigenstates becoming chaotic nearly simultaneously, 
a picture that differs from that predicted by KAM theory.

This may originate from the distinctive structure of integrability in many-body systems. In such systems there is 
no fixed number of conserved quantities; rather, their number typically increases with system size, 
while the range of quantum numbers associated with each conserved quantity is generally small 
and independent of the system size. The integrability of the XXZ chain is encoded in an extensive set of nonlocal 
or quasi-local conserved charges and in the organization of Bethe rapidities, rather than in invariant tori 
of a fixed-dimensional phase space. When the perturbation breaks these charges, it can couple a 
broad set of Bethe configurations. Consequently, a finite-order perturbative organization 
does not by itself imply the survival of a finite-density regular subsequence.

\section{Deformed Three-Mode Boson Model}
We have discussed and illustrated two elementary transitions from integrability to chaos, phaseout and morphing.
However, there are quantum systems whose transition is neither of them in pure form but a mixture. 
A deformed version of the three-mode boson model in Eq.~\eqref{eq:bsc-hamiltonian} is such a system. 

\subsection{Deformed Boson Model}
The deformed three-mode boson  model is described  by 
\begin{equation}
\hat H_{bd}(p,\lambda)
=
\hat H_{b}(p)
+\lambda\hat Q,
\label{eq:bsc-lambda-hamiltonian}
\end{equation}
where \(\hat H_{b}(p)\) is defined in
Eq.~\eqref{eq:bsc-hamiltonian}, and
\begin{equation}
\hat Q
=
\cos\!\left(\pi\hat n_0\right)
+
\frac{1}{2}\cos\!\left(\pi\hat m_q\right).
\label{eq:bsc-quasiperiodic-operator}
\end{equation}
The cosine functions are defined as functions of the number operators.
Thus, \(\hat Q\) is diagonal in the Fock basis
\(\lvert N_0,M_q\rangle\). It is also invariant under exchange of the
\(+\) and \(-\) modes, so the exchange-parity sectors remain exact.

At \(p=0\), the  Hamiltonian \(\hat H_{b}(p)\) and \(\hat Q\) are both
diagonal in this basis. The deformed model is therefore integrable for
every \(\lambda\) with spectrum
\begin{equation}
\begin{aligned}
E_{N_0,M_q}(0,\lambda)
={}&
\frac{c}{N}
\left[
N_0(N-N_0)+\gamma M_q^2
\right]
+\alpha N_0\\
&+\lambda\left[
\cos(\pi  N_0)
+\frac{1}{2}\cos(\pi  M_q)
\right].
\end{aligned}
\label{eq:bsc-lambda-integrable-spectrum}
\end{equation}
At finite \(p\), the coupling to mode \(0\) mixes Fock states with
different \(N_0\) and \(M_q\). The diagonal deformation changes their
relative energies while preserving exchange parity.

The size of this deformation has a different semiclassical scaling
from the hopping operator. Since
\(
\left\|\hat Q\right\|
\leq \frac{3}{2},
\)
its contribution to the energy per particle obeys
\begin{equation}
\left\|
\frac{\lambda\hat Q}{N}
\right\|
\leq
\frac{3|\lambda|}{2N}.
\label{eq:bsc-lambda-norm-bound}
\end{equation}
It therefore vanishes uniformly in the  classical energy per
particle as \(N\to\infty\).

This scaling can also be seen from the exact coherent-state
expectation, which is given by  
\begin{equation}
\begin{aligned}
q_N(n_0,m)
&=
\frac{1}{N}
\langle\bm z;N|\hat Q|\bm z;N\rangle\\
&=
\frac{3}{2N}
\operatorname{Re}
\left(1-2n_0\right)^N
\end{aligned}
\label{eq:bsc-quasiperiodic-symbol}
\end{equation}
Consequently, the deformation contributes
\(\lambda q_N(n,m)\) to the coherent-state energy per particle and does
not significantly modify the   classical Hamiltonian for \(N\gg 1\).

The vanishing of the deformation in the energy per particle does not
make it irrelevant to finite-size spectra. The operator
\(\lambda\hat Q\) produces bounded shifts in the total
energy, while individual level spacings in a fixed energy-density
window become small as the Hilbert-space dimension grows. The
alternating term can therefore modify finite-size level detunings and
eigenstates even though it leaves the  classical Hamiltonian
unchanged.

The role of this deformation is therefore more specific than a further
increase of classical chaos. At \(p=0\), the Hamiltonian remains integrable
for every \(\lambda\), whereas at finite \(p\) the question is whether the
normal-form states of the undeformed boson model remain spectrally isolated. We test this
by keeping the same energy window and comparing the Husimi projections,
energy-entropy structure, overlap weights, and extrapolated \(P(0)\) as
\(\lambda\) is increased. The relevant hypothesis is that a lattice-scale
term can reduce the quantum-classical state correspondence and the
independence of the regular and chaotic subsequences even when the global
value of \(\langle r\rangle\) changes only weakly. This is a controlled
numerical test, rather than a general statement that every non-smooth
perturbation destroys KAM phenomenology.

\subsection{Spectral Statistics}

\begin{figure}[htbp]
  \centering
  \begin{subfigure}{0.49\columnwidth}
    \centering
    \phantomcaption
    \stackinset{l}{1pt}{t}{1pt}{\footnotesize(\thesubfigure)}{%
      \includegraphics[width=\linewidth]{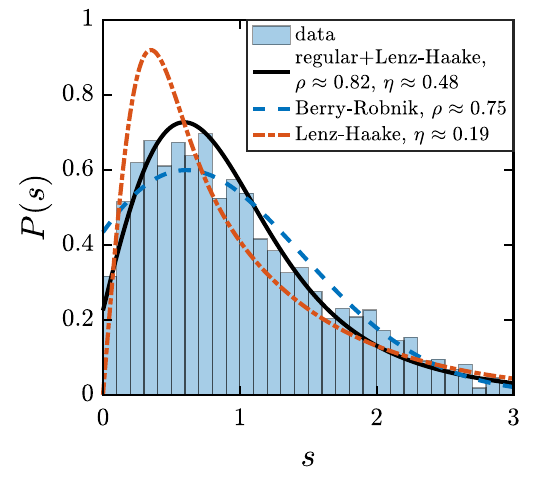}}
    \label{fig:lama}
  \end{subfigure}
  \hfill
  \begin{subfigure}{0.49\columnwidth}
    \centering
    \phantomcaption
    \stackinset{l}{1pt}{t}{1pt}{\footnotesize(\thesubfigure)}{%
      \includegraphics[width=\linewidth]{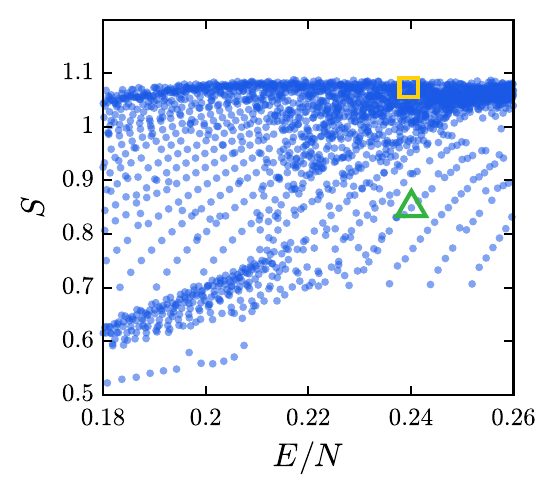}}
    \label{fig:lamb}
  \end{subfigure}

  \begin{subfigure}{0.49\columnwidth}
    \centering
    \phantomcaption
    \stackinset{l}{1pt}{t}{1pt}{\footnotesize(\thesubfigure)}{%
      \includegraphics[width=\linewidth]{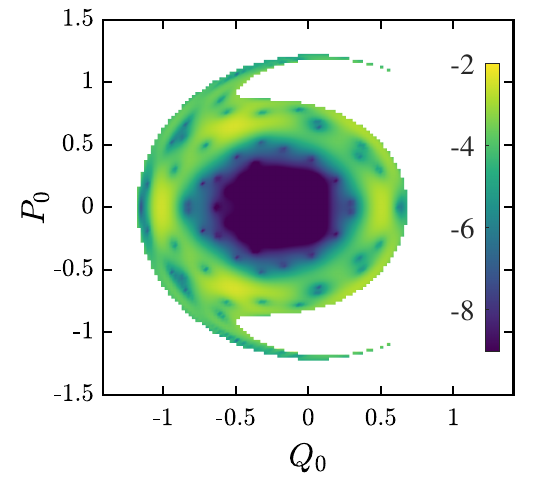}}
    \label{fig:lamc}
  \end{subfigure}
  \hfill
  \begin{subfigure}{0.49\columnwidth}
    \centering
    \phantomcaption
    \stackinset{l}{1pt}{t}{1pt}{\footnotesize(\thesubfigure)}{%
      \includegraphics[width=\linewidth]{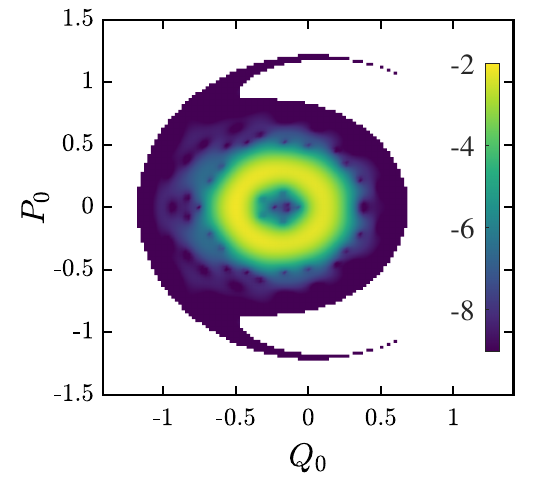}}
    \label{fig:lamd}
  \end{subfigure}
  \caption{(a) Level-spacing statistics for \(N=180\), \(\lambda=0.6\), and \(p=0.036\), with \(P(0)=0.25\). (b) Energy-entropy scatter plot at \(N=180\), \(p=0.055\), and \(\lambda=0.01\). One can see that the lattice-like structure in the low-entanglement region is gradually destroyed. (c) Husimi projection of the \(3193\)rd eigenstate onto the Poincar\'e section, corresponding to the yellow square in panel (b). (d) Husimi projection of the \(3177\)th eigenstate onto the Poincar\'e section, corresponding to the green triangle in panel (b).}
  \label{fig:lam}
\end{figure}

For each value of \(\lambda\), we determine the hopping strength \(p_*(\lambda)\) 
at which the deformed boson model reaches the transition point, defined by the same 
target value \(\langle r\rangle_*=0.46\) . 
We then estimate the intercept \(P(0)\) at this matched point using the short-spacing 
extrapolation described in Sec. \ref{sec:num}. 
This procedure allows us to compare spectra with similar global gap-ratio statistics while tracking whether 
their short-spacing behavior---and hence their organization into regular and chaotic 
subsequences---changes with the deformation. Fig. \ref{fig:lama} shows typical energy-level statistics in the presence of the deformation term.

At \(\lambda = 0.6\), we set \(p = 0.036\) to maintain \(\langle r \rangle = 0.46\). In this case, linear extrapolation yields \(P(0) = 0.25\), significantly below the value in the absence of the perturbation term. Meanwhile, the level-spacing statistics cannot be fitted by either the Berry--Robnik or the Lenz--Haake distribution alone; instead, a combination of integrable and Lenz--Haake-type spectra is required. 
It should also be noted that, in the classical limit, a smaller \(p\) is expected to increase the integrable fraction of the system. This implies that, 
while the degree of chaos of the system remains essentially unaltered, the residual integrable structure is destroyed, 
and the regular and chaotic eigenstates become mixed. From this perspective, the deformed boson model 
does not correspond to either of the two Hamiltonians in Sec.~\ref{theo}. The deformed boson model 
can instead be viewed as a residual regular subsequence embedded in a spectrum-wide evolving background. 
As \(\lambda\) increases, the decreasing \(P(0)\) is consistent with stronger coupling 
between the regular states and the evolving background, a reduced regular-state density, or both.

The explicit eigenstate calculations support this interpretation. Figure~\ref{fig:lamb} shows the energy-entropy scatter plot. 
In the low-entanglement region, the lattice-like pattern visible in the undeformed boson model becomes less distinct. 
Representative Husimi projections for the deformed model are shown in Figs.~\ref{fig:lamc} and \ref{fig:lamd}. 
It is clear from the figures that the Husimi distribution of the representative \(3177\)th eigenstate 
extends into the classically chaotic region rather than remaining concentrated near the regular island, 
while that of the representative \(3193\)rd eigenstate 
extends into the classically regular region. 
They show that a bounded, lattice-scale term can reorganize finite-\(N\) levels and eigenstates 
even when it leaves the leading classical limit unchanged. Since the \(p=0\) Hamiltonian remains 
diagonal for every \(\lambda\), the deformation also preserves the integrable endpoint. 
The above results show that, as the smooth quantum-classical correspondence is gradually destroyed, 
the residual integrable structure is likewise destroyed, even though the degree 
of chaos of the system remains essentially unchanged.

With the deformation term included, the extrapolated intercept decreases at the same mean gap ratio.
Together with the erosion of the Peres-lattice structure and the spreading of the Husimi distributions, 
this supports a loss of the regular-state organization found in the undeformed boson model. 
This further indicates that the smoothness of the quantum-classical correspondence determines whether 
a well-defined KAM structure exists in the quantum system. As the quantum-classical correspondence is gradually destroyed, 
the regular subsequences likewise disappear, and the KAM framework gradually ceases to apply. 
Because the deformation preserves the bosonic Hilbert space and the integrable \(p=0\) endpoint, 
it provides a more direct test of the role of smooth semiclassical structure than the comparison between 
the undeformed boson model and the XXZ spin chain alone. 

\section{Discussion and Conclusion}

The central result of this work is that the quantum integrable-to-chaotic transition 
comes in two elementary forms, and that a generic system may realize a mixture of them. In the phaseout transition, realized by the bosonic trimer, regular eigenstates remain anchored to the surviving structures of the classical phase space: they localize on regular islands, are accurately reproduced by the quantum normal form, form an approximately Poissonian subsequence statistically independent of the chaotic levels, and produce a nonzero intercept $P(0)$ that remains stable as $N$ grows. In the morphing transition, realized by the perturbed XXZ chain, level repulsion spreads through the whole resolved spectrum: $P(0)$ vanishes within uncertainty at all sizes studied, the spacing statistics follow the Lenz-Haake form, and no lattice-like family of low-entanglement states appears. The deformed trimer lies between these limits: a bounded lattice-scale perturbation, whose contribution to the classical energy per particle vanishes as $N\to\infty$, progressively erodes the Peres-lattice organization, delocalizes the Husimi distributions, and suppresses $P(0)$, even though the global degree of chaos measured by $\langle r\rangle$ is held fixed.

For quantum systems with $L$ subsystems of Hilbert space dimension $D_{\rm sub}$, 
we expect the phaseout transition to occur in systems with $D_{\rm sub}\gg 1$ and $L\sim 1$
and the morphing transition to happen in systems with $D_{\rm sub}\sim 1$ and $L\gg 1$. 
The former usually has a well defined classical limit at the semi-classical limit $D_{\rm sub}\rightarrow\infty$. 
For quantum systems with $D_{\rm sub}\sim 10$ and $L\sim 10$, we expect their transition
from integrable to chaotic is a mixture of phaseout and morphing with the energy level spacings 
obeying a distribution interpolating  the BR and LH.  It is highly desirable to find an analytic expression 
for such a distribution. 

We have used several diagnostic tools, such as entanglement and Husimi projection, 
to distinguish the two elementary transitions: phaseout and morphing. However, the most
essential tool is the spectral statistics, which can be used alone to characterize these two transitions.

In summary, we have shown that the quantum transition from integrability to chaos comes in two elementary forms. In the previously known phaseout (quantum KAM) transition, a finite-density family of regular eigenstates---organized by the surviving structures of the classical phase space---forms an approximately independent subsequence that gradually shrinks and disappears. In the morphing transition identified here, which has no classical counterpart, all eigenstates hybridize and turn chaotic concurrently, and no regular subsequence survives at any stage of the crossover. The two transitions are sharply distinguished by their spectral organization: an approximately independent superposition of regular and chaotic levels with Berry-Robnik statistics and a finite intercept $P(0)$, versus global level repulsion with Lenz-Haake statistics and $P(0)=0$. Our study of three models demonstrates that a bosonic trimer with a smooth semiclassical limit realizes the phaseout transition, that the perturbed XXZ chain realizes the morphing transition, and that a deformed trimer realizes a mixture in which the regular subsequence erodes even at a fixed global degree of chaos. This establishes the morphing transition as a second elementary paradigm of quantum integrability breaking, and shows that the phaseout (quantum KAM) scenario is conditional on a smooth finite-dimensional quantum-classical correspondence: the formal existence of a quantum normal form alone does not guarantee the survival of regular eigenstates.

\begin{acknowledgments}
This work was supported by the National Natural Science Foundation of China (92365202, 12475011, and 11921005), the National Key R\&D Program of China (2024YFA1409002), and the Shanghai Municipal Science and Technology Project (25LZ2601100). The authors acknowledge ChatGPT and Kimi for assistance with numerical calculations, analytical derivations, and manuscript preparation. All AI-assisted results were independently verified, and the authors take full responsibility for this work.
\end{acknowledgments}

\end{document}